
%
\input phyzzx
\catcode`@=11
%
%
\newtoks\UT
\newtoks\monthyear
\Pubnum={UT-\the\UT}
\UT={725}
\monthyear={October, 1995}
\def\p@bblock{\begingroup \tabskip=\hsize minus \hsize
    \baselineskip=1.5\ht\strutbox \topspace-2\baselineskip
    \halign to\hsize{\strut ##\hfil\tabskip=0pt\crcr
    \the\Pubnum\cr\the\monthyear\cr
    }\endgroup}
\def\bftitlestyle#1{\par\begingroup \titleparagraphs
    \iftwelv@\fourteenpoint\else\twelvepoint\fi
    \noindent {\bf #1}\par\endgroup}
\def\title#1{\vskip\frontpageskip \bftitlestyle{#1} \vskip\headskip}
%
%
\def\acknowledge{\par\penalty-100\medskip \spacecheck\sectionminspace
    \line{\hfil ACKNOWLEDGEMENTS\hfil}\nobreak\vskip\headskip}
%
%

%
\def\journal#1&#2(#3){\begingroup \let\journal=\dummyj@urnal
    \unskip, \sl #1\unskip~\bf\ignorespaces #2\rm
    (\afterassignment\j@ur \count255=#3) \endgroup\ignorespaces}
\def\andjournal#1&#2(#3){\begingroup \let\journal=\dummyj@urnal
    \sl #1\unskip~\bf\ignorespaces #2\rm
    (\afterassignment\j@ur \count255=#3) \endgroup\ignorespaces}
\def\andvol&#1(#2){\begingroup \let\journal=\dummyj@urnal
    \bf\ignorespaces #1\rm
    (\afterassignment\j@ur \count255=#2) \endgroup\ignorespaces}

\def\NP{Nucl.~Phys.}
\def\PL{Phys.~Lett.}
\def\PR{Phys.~Rev.}

\catcode`@=12
%


\titlepage

\title{On Effective Superpotentials in Supersymmetric Gauge Theories}

\author{Izawa {\twelverm Ken-Iti}
\foot{\rm JSPS Research Fellow.}}
\address{Department of Physics, University of Tokyo \break
                    Tokyo 113, Japan}

\abstract{Effective superpotentials
are considered which include glueball chiral superfields
among their arguments in supersymmetric gauge theories.
It is seen that the accommodation
of glueball superfields is suitable
for their complete determination.}

\endpage

\doublespace


\def\a{\alpha}

\def\L{\Lambda}
\def\F{\Phi}

\def\j{\journal}
\def\o{\over}


\REF\Shi{M.A.~Shifman and A.I.~Vainshtein \j \NP &B277 (86) 456;
         \andvol &B359 (91) 571; \nextline
         M.~Dine and Y.~Shirman \j \PR &D50 (94) 5389; \nextline
         S.P.~de Alwis, hep-th/9508053.}

\REF\Ven{G.~Veneziano and S.~Yankielowicz \j \PL &B113 (82) 231;
         \nextline G.M.~Shore \j \NP &B222 (83) 446; \nextline
         C.P.~Burgess, J.-P.~Derendinger, F.~Quevedo,
         and M.~Quir{\' o}s, hep-th/9505171.}

\REF\Tay{T.R.~Taylor, G.~Veneziano, and S.~Yankielowicz
         \j \NP &B218 (83) 493.}

\REF\Dav{A.C.~Davis, M.~Dine, and N.~Seiberg \j \PL &B125 (83) 487;
         \nextline
         I.~Affleck, M.~Dine, and N.~Seiberg \j \NP &B241 (84) 493;
         \andvol &B256 (85) 557.}

\REF\Ger{J.-M.~G{\' e}rard and J.~Weyers \j \PL &B146 (84) 411;
         \nextline
         D.~Amati, K.~Konishi, Y.~Meurice, G.C.~Rossi,
         and G.~Veneziano \j Phys.~Rep. &162 (88) 169.}

\REF\Sei{N.~Seiberg \j \PL &B318 (93) 469; \nextline
         K.~Intriligator and N.~Seiberg, hep-th/9509066.}

\REF\Seib{N.~Seiberg \j \PR &D49 (94) 6857.}

\REF\Int{K.~Intriligator, R.G.~Leigh, and N.~Seiberg
         \j \PR &D50 (94) 1092.}

\REF\Kap{V.~Kaplunovsky and J.~Louis \j \NP &B422 (94) 57; \nextline
         K.~Intriligator \j \PL &B336 (94) 409; \nextline
         K.~Intriligator and N.~Seiberg, hep-th/9506084.}

\sequentialequations

%
%

Symmetries provide exact information on effective actions
for field theories. In particular, supersymmetry
is powerful enough
to sometimes determine effective superpotentials
\refmark{\Shi}
even uniquely\rlap.
\refmark{\Ven, \cdots, \Kap}

In gauge theories, certain internal symmetries
turn out to be anomolous.
There have been two ways for dealing with anomalous symmetries
to determine effective
superpotentials in supersymmetric gauge theories.
One way given by Taylor, Veneziano, and Yankielowicz
\refmark{\Tay}
is to introduce glueball chiral superfields
\refmark{\Ven}
to realize anomalous transformation laws for effective actions.
The other way given by Intriligator, Leigh, and Seiberg
\refmark{\Int}
is to view anomalous symmetries as explicitly broken
and utilize them to derive selection rules
\refmark{\Sei}
for effective superpotentials.

In this paper, we take advantage of the former way
in determination of effective superpotentials.
Two simple examples are presented for concreteness.

Let us first consider
supersymmetric SU($N_c$) gauge theory with
$N_f = N_c + 1$ flavors of
quark chiral superfields $Q_i$ and ${\overline Q}^i$
in the fundamental representations
$N_c$ and ${\overline N}_c$, respectively,
where $i$ denotes the flavor index: $i = 1, \cdots, N_f$.
We put a mass term for the $N_f$-th flavor
$$
  W_{tree} = m Q_{N_f} {\overline Q}^{N_f}
 \eqn\TREE
$$
as a tree-level superpotential.

We take a glueball superfield
\refmark{\Ven}
$$
  S \sim W_\a W^\a,
 \eqn\GLUE
$$
where $W_\a$ is the field-strength chiral superfield for
the SU($N_c$) gauge multiplet,
in addition to the meson superfield $M_i^j$ and the baryon
superfields $B^i$, ${\overline B}_j$ as variables
\refmark{\Ger, \Seib}
to describe an effective superpotential $W_{eff}$.

Now we introduce a term
$$
  W_{dyn} = -S \ln [S^{-1} \L^{-2N_c + 1}
                    (\det M - B^i M_i^j {\overline B}_j)] - S,
 \eqn\DYN
$$
which satisfies all the symmetry constraints
including an anomalous one
in the theory
with the mass $m$ regarded as a spurion superfield\rlap.
\refmark{\Sei}
Here $\L$ denotes a dynamical scale of the gauge interaction.

Then we may write the effective superpotential as follows:
$$
  W_{eff} = W_{dyn}
           + S f \left( {m M_{N_f}^{N_f} \o S},
                 {B^i M_i^j {\overline B}_j \o \det M} \right),
 \eqn\EFF
$$
where $f$ is a holomorphic function to be determined.
This is because $W_{dyn}$ saturates the anomaly and hence $W_{eff}-W_{dyn}$
satisfies all the `classical symmetries' in the theory.

The term $W_{dyn}$ exclusively leads to
an effective superpotential
$$
  W'_{dyn} = \L^{-2N_c + 1}(B^i M_i^j {\overline B}_j - \det M),
 \eqn\LDYN
$$
when the superfield $S$ is integrated out.
Since this superpotential is appropriate
\refmark{\Seib}
for the massless limit with $N_f = N_c + 1$, we impose a condition
that $f = 0$ for $m = 0$.
Then we get
$$
  f(x, y) = x,
 \eqn\FUNC
$$
provided the perturbation theory works
in the asymptotic regime $\L \rightarrow 0$ (and $m \rightarrow 0$).
Thus we conclude
$$
  W_{eff} = W_{dyn} + m M_{N_f}^{N_f}.
 \eqn\FCON
$$
Note that the expression \LDYN, in contrast to \DYN, seems
inappropriate in considering
the asymptotic regime due to its singularity
for $\L \rightarrow 0$.

Let us turn to the next example.
We consider supersymmetric SU$(2)_1$ $\times$ SU$(2)_2$ gauge theory with
a chiral superfield $\F$ in a representation ({\bf 2}, {\bf 2}),
whose dynamical scales
are given by $\L_1$ and $\L_2$.
We take glueball superfields $S_1$ and $S_2$ corresponding to the gauge groups
SU$(2)_1$ and SU$(2)_2$ in addition to a gauge-singlet chiral superfield
$$
  X \sim \F \F
 \eqn\GSIN
$$
as variables
\refmark{\Int}
to describe an effective superpotential $W_{eff}$.

We introduce an anomaly-saturating term
$$
  W_{dyn} = S_1 \ln{\L_1^5 \o S_1 X} + S_2 \ln{\L_2^5 \o S_2 X}
           - S_1 - S_2 + S_1 F \left( S_2 \o S_1 \right),
 \eqn\NDYN
$$
where $F$ denotes a holomorphic function to be determined.

We can add a tree-level mass term for $\F$ with mass $m$ to obtain
an effective superpotential
$$
  W_{eff} = W_{dyn} + S_1 f \left( {mX \o S_1}, {S_2 \o S_1} \right),
 \eqn\NEFF
$$
where $f$ is a holomorphic function which satisfies
a condition that $f = 0$ for $m = 0$.
By means of the asymptotic limit $\L_1, \L_2 \rightarrow 0$
(and $m \rightarrow 0$),
we get
$$
  W_{eff} = W_{dyn} + mX.
 \eqn\NEEF
$$
Integrating out the superfield $X$, we obtain
$$
  W'_{eff} = S_1 \ln{m \L_1^5 \o S_1 (S_1 + S_2)}
             + S_2 \ln{m \L_2^5 \o S_2 (S_1 + S_2)} + S_1 F.
 \eqn\LEFF
$$

On the other hand,
the massive theory is expected to yield pure supersymmetric
SU$(2)_1$ $\times$ SU$(2)_2$ gauge theory as its low-energy limit,
whose effective superpotential is given by
\refmark{\Ven}
$$
  W_{pure} = S_1 \ln{{\L'}_1^6 \o S_1^2} + S_2 \ln{{\L'}_2^6 \o S_2^2},
 \eqn\LOW
$$
where $\L'_1$ and $\L'_2$ are dynamical scales for the pure gauge theory.
Comparing this expression with \LEFF,
we see
$$
  S_1 F = S_1 \ln{S_1 + S_2 \o S_1} + S_2 \ln{S_1 + S_2 \o S_2}.
 \eqn\LCON
$$

More generally,
the above approach confirms the linearity
\refmark{\Int, \Kap}
of an effective superpotential
$$
  W_{eff} = W_{dyn} + W_{tree},
 \eqn\GEN
$$
where $W_{dyn}$ is an anomaly-saturating term independent
of the couplings in a tree-level term $W_{tree}$.

In conclusion, we have seen that it is suitable for determination of effective
superpotentials in supersymmetric gauge theories to introduce
glueball superfields among their arguments from the beginning.

%
\acknowledge

We would like to thank
H.~Murayama, and T.~Yanagida for valuable discussions.


\endpage

\refout

\bye